\documentclass[epj]{svjour}
\usepackage{pifont}
\usepackage{txfonts}
\usepackage{graphics}

\begin{document}

\title{Criticality of the Mean-Field Spin-Boson Model: Boson State Truncation and Its Scaling Analysis}

\author{Yan-Hua Hou\inst{1} \and Ning-Hua Tong\inst{1,}
\thanks{\emph{} e-mail: nhtong@ruc.edu.cn}
} \institute{Department of Physics, Renmin University of China,
Beijing 100872, P. R. China}

\date{Received: date / Revised version: date}

\abstract{The spin-boson model has nontrivial quantum phase
transitions at zero temperature induced by the spin-boson coupling.
The bosonic numerical renormalization group (BNRG) study of the
critical exponents $\beta$ and $\delta$ of this model is hampered by
the effects of boson Hilbert space truncation. Here we analyze the
mean-field spin boson model to figure out the scaling behavior of
magnetization under the cutoff of boson states $N_{b}$. We find that
the truncation is a strong relevant operator with respect to the
Gaussian fixed point in $0<s<1/2$ and incurs the deviation of the
exponents from the classical values. The magnetization at zero bias
near the critical point is described by a generalized homogeneous
function (GHF) of two variables $\tau=\alpha-\alpha_{c}$ and
$x=1/N_{b}$. The universal function has a double-power form and the
powers are obtained analytically as well as numerically. Similarly,
$m(\alpha=\alpha_{c})$ is found to be a GHF of $\epsilon$ and $x$.
In the regime $s>1/2$, the truncation produces no effect.
Implications of these findings to the BNRG study are discussed.
}

\authorrunning {Y.-H. Hou \and  N.-H. Tong}
\titlerunning{Criticality and scaling analysis of the Mean Field SBM induced by truncation} \maketitle

\section{Introduction}
   The spin-boson model (SBM) describing a quantum two-level system
coupled to a dissipative environment appears in many areas of
condensed matter physics \cite{RefQD:1,RefLC:2}. It is a simple
model for studying the dissipation and decoherence in a quantum
system subjected to interactions with environment, such as a qubit
for the quantum computations \cite{RefVP:3}. The rich
environment-induced quantum phase transitions in the spin-boson
model also attract much research attention recently. Experimentally,
an environment-induced localization transition has been observed in
the Josephson junction systems \cite{RefPP:4}, although typical
solid-state two-level systems have a coupling strength below the
critical threshold \cite{RefHF:5}. Ref.\cite{RefPF:6} showed that
optical forces allow realizing a variety of spin-boson models,
depending on the crystal geometry and the laser configuration. In
most realistic situations the boson bath spectrum is super-Ohmic
$(s>1)$ or Ohmic $(s=1)$. Recently an experimental set up for
realizing the spin-boson model in the sub-Ohmic $(0<s<1)$ and strong
coupling regime is proposed in the mesoscopic metal ring systems
\cite{RefTV:7}. ($s$ stands for the exponent of the bath spectral
function).

Many theoretical studies focus on the quantum phase transitions in
the spin-boson model \cite{RefLC:2}. For the super-Ohmic
dissipation, the system is always delocalized and there is no phase
transition. For the Ohmic dissipation, a quantum Kosterlitz-Thouless
(KT) transition separates the delocalized phase at small dissipation
from the localized phase at large dissipation, with critical
coupling strength $\alpha \approx 1$ in the small tunneling limit.
The study of the SBM with sub-Ohmic bath is more difficult. The path
integral with noninteracting blip approximation and adiabatic
renormalization found no transitions in this case \cite{RefLC:2}. In
contrast, the quantum-to-classical mapping theory for SBM showed
that a quantum phase transition existed for $0<s\leq1$
\cite{RefQD:1,RefLC:2,RefFC:8} and that the phase transition between
the localized and delocalized phases should be in the mean-field
universality class in the regime $0<s<1/2$ \cite{RefME:9,RefEH:10}.
Some important properties of the equilibrium and non-equilibrium
spin-boson model has been obtained from the flow equation method
\cite{RefKA:11,RefSM:12,RefTS:13}, from which a first order phase
transition was argued to exist in the sub-Ohmic case. Recently, a
perturbation approach based on unitary transformation was used to
study the dynamics of SBM with sub-Ohmic bath and the results also
suggested a first order localized-delocalized phase
transition\cite{RefLZ:14}.

    A comprehensive study on the critical behavior of SBM
for the sub-Ohmic dissipation begins when the non-perturbative
numerical renormalization group (NRG) method is extended to solve
boson problems \cite{RefBT:15,RefVT:16,RefBL:17}. A continuous phase
transition is disclosed and the critical exponents obtained by NRG
are non-classical and fulfill hyper-scaling relations. In the
$0<s<1/2$ regime, the NRG results for the critical exponents $\beta$
and $\delta$ are different from the classical ones ($\beta=1/2$ and
$\delta=3$). This is in contrast to the claim of the
quantum-to-classical mapping above \cite{RefME:9,RefEH:10}, hence
questioning the legitimacy of the mapping theory in this model.
Recently, the continuous time Monte Carlo calculation was applied to
the sub-Ohmic SBM \cite{RefWR:18}. It's found that in the regime
$0<s<1/2$ the critical exponents do take the classical values. A
sparse polynomial space approach (SPSA) together with a standard
exact diagonalization algorithm has been employed to calculate the
critical behaviors of the sub-Ohmic SBM \cite{RefAAH:19}. The
results confirmed the Gaussian critical fixed point in the regime
$0<s<1/2$. It is now realized that the exponents $\beta$ and
$\delta$ produced by NRG are incorrect in the regime $0<s<1/2$, due
to the boson state truncation used in the algorithm. How to avoid
this error in the bosonic NRG is still an open question.

The issue of quantum-to-classical mapping also appears in the
Bose-Fermi-Kondo model (BFKM). In the spin-isotropic case, S.
Kirchner \emph{et al.} adopted a dynamical large-$N$ limit of the
SU($N$) method to study the effect of the Berry phase term of the
spin path integral on the quantum critical properties. They
attributed the emergence of the interacting fixed point to the
interference of the Berry phase \cite{RefSQ:20}. In their recent
work, S. Kirchner \emph{et al.} argued again that the presence of
Berry phase changed the critical properties and claimed that the
mapping theory failed for the sub-Ohmic BFKM \cite{RefSQD:21}. A
scaling analysis for the Ising-BFKM was performed in
Ref.\cite{RefSQK:22}, which was believed to have the same critical
properties as the spin-boson model. The authors restated the failure
of the quantum to classical mapping for the QCP of the Ising-BFKM.
They argued that the continuum limit taken in the mapping failed to
capture the topological effect encoded in the Kondo spin-flips which
was essential to the nature of the QCP.

Local boson Hilbert space truncation is frequently used in the
numerical studies of boson systems, such as exact diagonalization
(ED), NRG and ED+DMFT study of the Bose-Hubbard model
\cite{RefWT:23}. In these algorithms, calculating with a finite
boson state number $N_{b}$ and then extrapolating the results to
$N_{b}=\infty$ is believed to be sufficient to yield correct
results. However, the NRG study of the spin-boson model in $0<s<1/2$
regime shows that the simple extrapolation cannot guarantee the
correctness of the critical exponents $\beta$ and $\delta$
\cite{RefVTB:24}. In order to figure out the role of $N_{b}$ in the
critical behavior of the order parameter $m$, in this paper, we
present a numerical analysis to the mean-field spin-boson model.
This model has a Gaussian critical fixed point, the same as the
spin-boson model in the regime $0<s<1/2$, and its exact critical
exponents are known, i.e., $\beta=1/2$ and $\delta=3$.

The structure of this paper is as follows. In Section $2$, we
introduce the mean-field spin-boson model and its star-form. The
methods of solution in the full as well as the truncated Hilbert
space are presented. In Section $3$, we present numerical results
for the critical exponents $\beta$, $\delta$, and $\gamma$. Scaling
analysis of the magnetization function is employed to interpret
these numerical results. The relations between our findings here and
the NRG study of the spin-boson model are discussed. In Section $4$
we end with a brief summary.

\section{Model and Method}
\label{sec:1}
\subsection{The mean-field spin-boson hamiltonian }
   The Hamiltonian of the spin-boson model reads $(\hbar=1)$
\cite{RefQD:1,RefLC:2}

\begin{eqnarray}\label{eq:1}
H&=&-\frac{\Delta}{2}\sigma_{x}+\frac{\epsilon
}{2}\sigma_{z}+\sum_{i}\omega_{i}a_{i}^{\dag}a_{i}+\frac{\sigma_{z}}{2}\sum_{i}\lambda_{i}(a_{i}^{\dag}+a_{i}).
\end{eqnarray}

Here, the Pauli matrices $ \sigma_{x}$ and $\sigma_{z}$ describe a
two-state system. $\epsilon $ is the energy difference and $\Delta$
is the tunneling strength between the two states. The environment is
modeled as a collection of harmonic oscillators, which serve as the
origin of dissipation\cite{RefQD:1,RefLC:2}. $a_{i}^{\dag}$ and
$a_{i}$ are creation and annihilation operators for the i-$th$
phonon mode with frequency $\omega_{i}$. $\lambda_{i}$ represents
the coupling between the two-state system and the i-$th$ phonon
mode. The effect of the harmonic environment is characterized by the
bath spectral function

\begin{eqnarray}\label{eq:2}
J(\omega)&=&\pi\sum_{i}\lambda_{i}^{2}\delta(\omega-\omega_{i}),
\end{eqnarray}
which completely determines the influence of environment on the two
level system. For simplicity, we use a power form of the spectral
function with an energy cutoff $\omega_{c}$

\begin{eqnarray}\label{eq:3}
J(\omega)&=&2\pi\alpha\omega_{c}^{1-s}\omega^{s}\Theta(\omega_c-\omega).
\end{eqnarray}
Here $\omega_{c}=1$ is the energy unit. $\alpha$ is a dimensionless
coupling constant that characterizes the dissipation strength. The
index $s$ accounts for certain physical environment to which the
two-state system couples.

At zero temperature and zero bias, the competition between the
quantum mechanical tunneling of the two states (leading to a
delocalized phase) and the effect of the spin-bath coupling (
tending to localize the system into spin up or spin down state)
leads to a phase transition between the two phases at a critical
coupling $\alpha_{c}$ (for a fixed $\Delta$).

   We focus on the mean field spin-boson model. It is exactly solvable in the
infinite boson Hilbert space, with an analytical expression of
critical coupling $\alpha_c$ and the classical exponents. In the
truncated boson Hilbert space, it can be solved numerically at high
precision. It hence enables us to focus on the effect of the local
boson Hilbert space truncation on the critical behavior. Carrying
out the mean-field approximation to the spin-boson model, we obtain
the mean-field Hamiltonian

\begin{eqnarray}\label{eq:4}
H_{MF}&=&-\frac{\Delta}{2}\sigma_{x}+\frac{\epsilon
}{2}\sigma_{z}+\sum_{i}\omega_{i}a_{i}^{\dag}a_{i}
\nonumber \\
   &+& \frac{\langle \sigma_z\rangle}{2} \sum_{i} \lambda_i \left(a_{i}^{\dagger}+a_{i}
   \right) + \frac{\sigma_z}{2} \sum_{i} \lambda_i \langle a_{i}^{\dagger}+a_{i}
   \rangle  \nonumber \\
   &-& \frac{\langle \sigma_z\rangle}{2}\sum_{i} \lambda_i \langle
   a_{i}^{\dagger}+a_{i}.
   \rangle.
\end{eqnarray}
Neglecting the constant term, it can be written as the sum of
decoupled Hamiltonians for isolated spin and displaced free bosons:

\begin{equation}\label{eq:5}
  H_{MF}=H_{spin}+H_{boson},
\end{equation}
with

\begin{equation}\label{eq:6}
   H_{spin}=-\frac{\Delta}{2}\sigma_{x}+ \left[ \frac{\epsilon }{2}+ \sum_{i} \frac{\lambda_i}{2} \langle
   a_{i}^{\dagger}+a_{i}
   \rangle \right] \sigma_{z},
\end{equation}
and

\begin{equation}\label{eq:7}
   H_{boson}=\sum_{i}\omega_{i}a_{i}^{\dag}a_{i}+ \frac{\langle \sigma_z\rangle}{2} \sum_{i} \lambda_i \left(a_{i}^{\dagger}+a_{i}
   \right).
\end{equation}

   In the truncated boson Hilbert space, $a_{i}$ and $a_{i}^{\dagger}$
are no longer canonical boson operators. One needs to resort to
numerical calculations to solve Eq.(\ref{eq:5})-(\ref{eq:7}). Then,
one has to specify the form of the SBM Hamiltonian, i.e., to
parameterize the environment spectrum $J(\omega)$ and assign values
to parameters $\lambda_i$ and $\omega_i$. One common way to
discretize and parameterize the bath degrees of freedom used in the
NRG calculation is the logarithmic discretization. In order to make
connection to the NRG studies, we use the same parametrization as in
NRG. Following the procedure in Ref.\cite{RefBL:17} with an
additional mean-field approximation we arrive at the star-form
mean-field Hamiltonian below

\begin{equation}\label{eq:8}
  H_{MF}^{star}=H_{spin}^{star}+H_{boson}^{star},
\end{equation}
with

\begin{equation}\label{eq:9}
   H_{spin}^{star}=-\frac{\Delta}{2}\sigma_{x}+ \left[ \frac{\epsilon}{2}+ \frac{1}{2\sqrt{\pi}} \sum_{n}\gamma_{n}
   \langle a_{n}+a_{n}^{\dag} \rangle \right] \sigma_{z},
\end{equation}
and

\begin{equation}\label{eq:10}
   H_{boson}^{star}=\sum_{n}\xi_{n}a_{n}^{\dag}a_{n} + \frac{\langle \sigma_z\rangle}{2\sqrt{\pi}}
   \sum_{n}\gamma_{n}\left(a_{n}+a_{n}^{\dag} \right).
\end{equation}
Here, the logarithmic discretization gives

\begin{eqnarray}\label{eq:11}
\gamma_{n}^{2}=\frac{2\pi\alpha}{1+s}\left[1-\Lambda^{-(1+s)}\right]
\Lambda^{-n(1+s)}\omega_{c}^{2},
\end{eqnarray}
and
\begin{eqnarray}\label{eq:12}
\xi_{n}=\frac{1+s}{2+s}\frac{1-\Lambda^{-(2+s)}}{1-\Lambda^{-(1+s)}}\Lambda^{-n}\omega_{c}.
\end{eqnarray}
$\Lambda >1$ is the logarithmic discretization parameter.\\
\subsection {Numerical methods for truncated $H_{MF}^{star}$ }
   In the truncated boson Hilbert space, due to the decoupling of boson
modes in $H_{MF}^{star}$, it is possible to solve the boson part
$H_{boson}^{star}$ by exact diagonalization for each truncated mode.
The obtained boson average $\langle a_{n}^{\dagger} + a_{n} \rangle$
for $n=0, 1, ..., N_{c}$ are input into the spin Hamiltonian to
solve for $\langle \sigma_z \rangle$. This iteration continues until
convergence is reached.

   For the calculation of susceptibility $\chi$, we start from the
self-consistency equation of the order parameter $m=\langle \sigma_z
\rangle$ at zero temperature,

\begin{equation}\label{eq:13}
m(\epsilon, t,
\Delta)=\frac{\Delta^{2}}{\Delta^{2}+(\epsilon+t)^2+(\epsilon+t)\sqrt{(\epsilon+t)^2+\Delta^{2}}}-1,
\end{equation}
with

\begin{eqnarray}\label{eq:14}
t\equiv
t(\epsilon)&=&\frac{1}{\sqrt{\pi}}\sum_{n=0}^{N_{c}}\gamma_{n}\overline{a}_{n},
\end{eqnarray}
where

\begin{eqnarray}\label{eq:15}
\overline{a}_{n}&=&<a_{n}+a_{n}^{\dag}>,
\end{eqnarray}
and $N_{c}$ boson modes are used. After some algebra, we obtain the
final equation of the susceptibility

\begin{eqnarray}\label{eq:16}
\chi^{-1}&=&-\frac{\left(t_{0}^{2}+\Delta^{2}+t_{0}\sqrt{t_{0}^{2}
+\Delta^{2}}\right)^{2}\sqrt{t_{0}^{2}+\Delta^{2}}}
{\Delta^{2}\left(2t_{0}\sqrt{t_{0}^{2}+\Delta^{2}}
+2t_{0}^{2}+\Delta^{2} \right)}\nonumber\\
& &
+\frac{1}{\pi}\sum_{n=0}^{N_{c}}\sum_{k=2}^{N_{b}}\gamma_{n}^{2}\frac{<
g_{n}|a_{n}+a_{n}^\dag|k_{n}>^{2}}{\epsilon _{n}^{k}-\epsilon
_{n}^{g}}.
\end{eqnarray}
Here, $t_{0}$ is the $t(\epsilon)$ in Eq.(\ref{eq:14}) at zero
external bias. $|g_{n}>$, $\epsilon _{n}^{g}$ and $|k_{n}>$,
$\epsilon _{n}^{k}$ denote the eigenvectors and eigenvalues of the
ground state and the $k_{th}$ excited state of the $n_{th}$ boson
mode, respectively. Due to the Hilbert space truncation, the
summation of $k$ is limited to $N_{b}$ for each mode $n$.
Eq.(\ref{eq:16}) is evaluated numerically after the iterative
solution is converged.

   We use total $101$ different boson modes in the calculation, i.e.,
the summation of boson modes is cut off at $N_{c}=100$. Due to the
exponential decay of $\gamma_n$ and $\xi_{n}$, this summation is
already numerically exact. For each boson mode, we retain $N_{b}$
states. For simplicity, we use the boson number eigen states $|0>$,
$|1>$, ..., $|N_{b}-1>$ as bases. In our calculation, the NRG
parameter $\Lambda=2.0$ is used.
\section{Results and Discussions}
\subsection {Exact solution at $N_{b}=\infty$}
In the full Hilbert space, $a_{i}$ and $a_{i}^{\dagger}$ are
canonical boson operators obeying the common commutation relation
$\left[a_{i}, a_{j}^{\dagger} \right]= \delta_{ij}$. One gets the
self-consistent mean-field equations for $H_{MF}$

\begin{eqnarray}\label{eq:17}
m&=&\frac{m\Gamma-\epsilon }{2\lambda}\tanh(\beta \lambda),
\end{eqnarray}
and

\begin{eqnarray}\label{eq:18}
\lambda&=&\frac{1}{2}\sqrt{(m\Gamma-\epsilon)^{2}+\Delta^{2}},
\end{eqnarray}
with

\begin{eqnarray}\label{eq:19}
\Gamma&=\int_{0}^{\infty}\frac{J(\omega)}{\pi\omega}d\omega=2\alpha\omega_{c}/s.
\end{eqnarray}
This set of self-consistent equations can be solved and the critical
coupling strength for a given temperature $T$ is

\begin{eqnarray}\label{eq:20}
\alpha_{c}&=&\frac{s\Delta}{2 \tanh( \beta \Delta /2 )\omega_{c}}.
\end{eqnarray}
It reduces to $\alpha_{c}=s\Delta/(2\omega_{c})$ at $T=0$.

To study the quantum phase transition at $T=0$, we focus on the
following critical exponents $\beta$, $\delta$, and $\gamma$, that
are related to the behavior of the order parameter $m$
\cite{RefVT:16},

\begin{eqnarray}\label{eq:21}
  m(\alpha>\alpha_{c}, T=0, \epsilon=0)\propto(\alpha-\alpha_{c})^{\beta},
\end{eqnarray}

\begin{eqnarray}\label{eq:22}
m(\alpha=\alpha_{c}, T=0, \epsilon)\propto|\epsilon|^{1/\delta},
\end{eqnarray}

\begin{eqnarray}\label{eq:23}
\chi_{loc}(\alpha<\alpha_{c}, T=0,
\epsilon=0)\propto(\alpha_{c}-\alpha)^{-\gamma}.
\end{eqnarray}
The classical values $\beta=1/2$ and $\delta=3$ are obtained, as it
should be. Via the partial derivative of m with respect to the local
external field $\epsilon$, we get the zero temperature
susceptibility

\begin{eqnarray}\label{eq:24}
\chi= \left\{
\begin{array}{lll}
\frac{s}{2\omega_{c}(\alpha-\alpha_{c})},\,\,\,\,\,\,\,\,\,    & (\textrm{$\alpha<\alpha_{c}$});\\
 &\\
 \frac{s}{4\omega_{c}(\alpha_{c}-\alpha)},\,\,\,\,\,\,\,\,\,     &
(\textrm{$\alpha>\alpha_{c}$}),
\end{array} \right.
\end{eqnarray}
which gives $\gamma=1$.

   Note that for the full boson Hilbert space, due to the linear nature
of the logarithmic discretization and the transformation, the two
forms of mean-field Hamiltonian, $H_{MF}$ and $H_{MF}^{star}$ are
essentially equivalent and belong to the same universality class.
The mean-field approximation and the transformation can be
interchanged in sequence. Therefore, the self-consistent mean-field
equations Eq.(\ref{eq:17}-\ref{eq:18}) still hold for
$H_{MF}^{star}$, but with $\Gamma$ replaced by

\begin{eqnarray}\label{eq:25}
\Gamma&=&\sum_{n=0}^{\infty}\frac{\gamma_{n}^{2}}{\pi\xi_{n}}\nonumber\\
 &=&\frac{2\alpha\omega_{c}(s+2)[1-\Lambda^{-(s+1)}]^{2}}{(1+s)^{2}(1-\Lambda^{-s})[1-\Lambda^{-(s+1)}]}.
\end{eqnarray}
The zero temperature critical coupling $\alpha_{c}$ now reads

\begin{eqnarray}\label{eq:26}
\alpha_{c}&=&\frac{\Delta(s+1)^{2}(1-\Lambda^{-s})[1-(\Lambda^{-(s+2)})]}{2\omega_{c}(s+2)[1-(\Lambda^{-(s+1)})]^{2}}.
\end{eqnarray}
In the limit $\Lambda \rightarrow 1$, $\alpha_{c}$ tends to the
critical value $\alpha_c(T=0)=s\Delta/(2\omega_{c})$, being
consistent with Eq.(\ref{eq:20}).

\subsection {Analytical solution at $N_{b}=2$}
   We obtained the critical exponents $\beta$ and $\delta$ for $H_{MF}^{star}$
analytically in the special case at $N_{b}=2$. (See in the Appendix
for details.) We found that
$\alpha_{c}(N_{b}=2)=\alpha_{c}(N_{b}=\infty)$ as given in
Eq.(\ref{eq:26}), suggesting that $\alpha_{c}$ is independent of the
truncation $N_{b}$. This has indeed been observed in our numerical
calculations at various $N_{b}$'s. We also found for $N_{b}=2$

\begin{eqnarray}\label{eq:27}
\beta&=&\left\{
\begin{array}{lll}
\frac{1}{2},\,\,\,\,\,\,\,\,\, & (\textrm{$s\geq\frac{1}{2}$});\\
 &\\
 \frac{1-s}{2s},\,\,\,\,\,\,\,\,\,&
(\textrm{$0<s<\frac{1}{2}$}).
\end{array} \right.
\end{eqnarray}

In the case of $s=0$, we get $\alpha_{c}=0$. $\beta$ diverges and
the magnetization has the form:

\begin{eqnarray}\label{eq:28}
m\propto\alpha^{-\frac{1}{2}}exp[-\frac{\Delta
ln\Lambda(1+\Lambda^{^{-1}})}{8\alpha\omega_{c}(1-\Lambda^{-1})}].
\end{eqnarray}

\begin{figure}
\resizebox{0.52\textwidth}{!}{
  \includegraphics{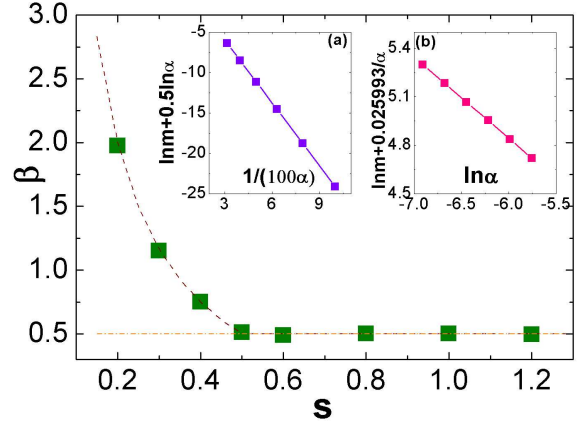}}
\caption{ Exponent $\beta$ as a function of $s$ at the parameters:
$\Delta=0.1$, $\epsilon=0$, $N_{b}=2$. Dash line represents the
analytical data; Olive squares symbol numerical data; Dash-dot line
is for $N_{b}=\infty$. The inset shows the magnetization at $s=0$ as
a function of $\alpha$ with $N_{c}=2000$. }\label{fig:1}
\end{figure}

The numerical results agree well with these results, as revealed in
Fig. \ref{fig:1}. The special case for $s=0$ is manifested as the
inset in Fig. \ref{fig:1}. In Fig. (1.a), the slope is approximately
-0.02599 for $\Lambda=2.0$, being consistent with the power of the
exponential part in Eq.(\ref{eq:28}). In Fig. (1.b), the slope is
-0.5.

\begin{figure}
\resizebox{0.52\textwidth}{!}{
  \includegraphics{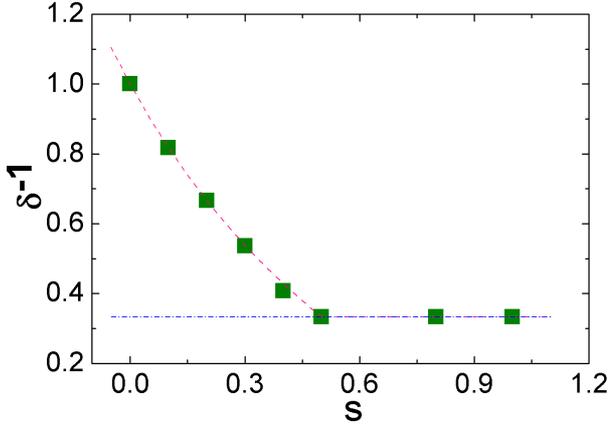}} \caption{Exponent $\delta^{-1}$ as a function of $s$ at the
parameters: $\Delta=0.1$, $\alpha=\alpha_{c}$, $N_{b}=2$. The dash
line is the analytical result; The olive squares are numerical data;
The dash-dot line is for $N_{b}=\infty$.}\label{fig:2}
\end{figure}

   Similarly, both analytical and numerical results for the critical
exponent $\delta$ are available at $N_{b}=2$, as shown in Fig.
\ref{fig:2}.
\begin{eqnarray}\label{eq:29}
\delta&=&\left\{
\begin{array}{lll}
3,\,\,\,\,\,\,\,\,\, & (\textrm{$s\geq\frac{1}{2}$});\\
 &\\
 \frac{1+s}{1-s},\,\,\,\,\,\,\,\,\,&
(\textrm{$0<s<\frac{1}{2}$}).
\end{array} \right.
\end{eqnarray}

It is noted that in $0<s<1/2$, both $\beta$ and $\delta$ for
$N_{b}=2$ agree with the corresponding exponents in the spin-boson
model obtained from NRG. A natural question is how the nonclassical
critical exponents $\beta$ and $\delta$ in the regime $0<s<1/2$ as
given in Eq.(\ref{eq:27}) and Eq.(\ref{eq:29}) at $N_{b}=2$ change
to classical ones at $N_{b}=\infty$. In the following section, we
explore this issue at intermediate $N_{b}$'s.

\subsection {Numerical analysis at intermediate $N_{b}$'s}
\begin{figure}
\resizebox{0.52\textwidth}{!}{
  \includegraphics{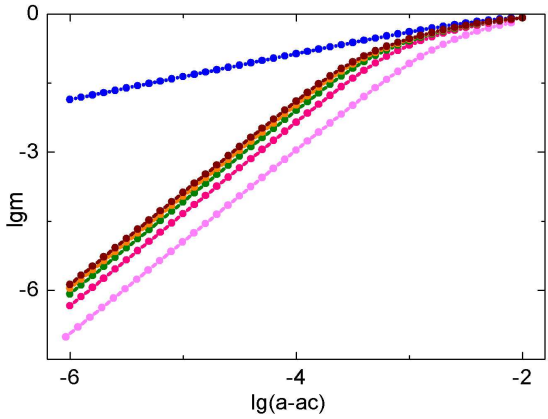}} \caption{lg$m$ v.s. lg$\tau$
  for different truncations at the
parameters: $\Delta=0.1$, s=0.2, $\epsilon$=0.0. The truncations are
$N_{b}$=2, 5, 10, 15, 20 and $\infty$ from bottom to top.}
\label{fig:3}
\end{figure}

At intermediate $N_{b}$'s, we calculate $<a_{n}+a_{n}^{\dag}>$ by
exact diagonalization and solve the mean-field equation iteratively.
For $s\geq1/2$, the numerical results always yield the classical
critical exponents $\beta=1/2$ and $\delta=3$, irrespective of the
boson state truncation $N_{b}$'s. But for $0<s<1/2$, this is no
longer the case. In Fig.\ref{fig:3}, we plot the dependence of the
average magnetization $m$ on the dissipation strength $\alpha$ for
$s=0.2$, at different $N_{b}$'s ($2\leq N_{b}\leq\infty$).
$\alpha_{c}$ is calculated from Eq.(\ref{eq:26}) at given
parameters. It is clearly seen that in the small $\alpha-\alpha_{c}$
limit, $m$ exhibits a perfect power law and the slope $\beta_{d}$ is
identical to that of $N_{b}=2$, namely $\beta_{d}=(1-s)/(2s)$,
deviating from the classical exponent dramatically. Only for
$N_{b}\rightarrow\infty$ do we recover the anticipated mean-field
result. In the upper region of the curve, the magnetization at
finite but large $N_{b}$ tends to overlap with that at
$N_{b}=\infty$, giving a different slope $\beta_{u}=1/2$. As a
result, two different power laws appear in the lower and upper
regimes of the curve. In the following, we carry out a scaling
analysis for $m$ data with respect to parameters
$\tau=\alpha-\alpha_{c}$ and $x=1/N_{b}$ on the basis of the
generalized homogeneous function (setting $\epsilon=0$)
\cite{RefAH:25}.

   We assume that the singular part of $m$ is a GHF of $\tau$ and $x$.
That is,

\begin{eqnarray}\label{eq:30}
m(\lambda^{a_{\tau}}\tau, \lambda^{a_{x}}x)&=&
\lambda^{a_{m}}m(\tau, x).
\end{eqnarray}

Here $\lambda$ is a positive number. Taking a logarithmic form of
the equation above, we arrive at the following:

\begin{eqnarray}\label{eq:31}
\overline{m}(\overline{\tau}+a_{\tau}\overline{\lambda},\overline{x}+a_{x}\overline{\lambda})
&=&\overline{m}(\overline{\tau},\overline{x})+a_{m}\overline{\lambda},
\end{eqnarray}
where $\overline{i}$=lg$i$, $i=m$, $\tau$, $x$, $\lambda$. This
equation implies that if $\overline{m}$, $\overline{\tau}$ and
$\overline{x}$ are shifted by $\Delta\overline{m}$,
$\Delta\overline{\tau}$, $\Delta\overline{x}$, respectively, the
curves will collapse. The ratios of the shifts give the critical
exponents $\Delta\overline{m}/\Delta \overline{\tau}=a_{m}/a_{\tau}$
and $\Delta\overline{m}/\Delta \overline{x}=a_{m}/a_{x}$. Assigning
$\overline{\lambda}=-\overline{x}/a_{x}$ in Eq.(\ref{eq:31}), we get

\begin{eqnarray}\label{eq:32}
\overline{m}(\overline{\tau}-\frac{a_{\tau}}{a_{x}}\overline{x},
0)&=& \overline{m}(\overline{\tau},
\overline{x})-\frac{a_{m}}{a_{x}}\overline{x}.
\end{eqnarray}

For simplicity, we denote
$z=\overline{\tau}-\frac{a_{\tau}}{a_{x}}\overline{x}$. Fig.
\ref{fig:3} suggests that we can assume a universal function as:

\begin{eqnarray}\label{eq:33}
\overline{m}(\overline{\tau}-\frac{a_{\tau}}{a_{x}}\overline{x}, 0)=
\overline{m}(z, 0)= \left\{
\begin{array}{lll}
\beta_{u}z+C_{u},\,\,\,\,\,\,& (z\gg
z_{0});\\
&\\
\beta_{d}z+C_{d},\,\,\,\,\,\,     & (z\ll z_{0}).
\end{array} \right.
\end{eqnarray}
$z_{0}$ is some crossover value separating two regimes with
different power laws. This gives

\begin{eqnarray}\label{eq:34}
\overline{m}(\overline{\tau},\overline{x})= \left\{
\begin{array}{lll}
\beta_{u}\overline{\tau}+\overline{x}(\frac{a_{m}}{a_{x}}-\frac{a_{\tau}}{a_{x}}\beta_{u})+C_{u},\,\,\,
& (\tau^{\frac{1}{a_{\tau}}}\gg
cx^{\frac{1}{a_{x}}});\\
&\\
\beta_{d}\overline{\tau}+\overline{x}
(\frac{a_{m}}{a_{x}}-\frac{a_{\tau}}{a_{x}}\beta_{d})+C_{d},\,\,\, &
(\tau^{\frac{1}{a_{\tau}}}\ll cx^{\frac{1}{a_{x}}}).
\end{array} \right.
\end{eqnarray}
$C_{u}$, $C_{d}$ and $c$ are constants. The truncation independence
of the upper power in Fig. \ref{fig:3} requires that the coefficient
of $\overline{x}$ in $\tau^{\frac{1}{a_{\tau}}}\gg
cx^{\frac{1}{a_{x}}}$ regime should be zero, namely
$a_{m}/a_{x}=\beta_{u}a_{\tau}/a_{x}$ or $\beta_{u}=a_{m}/a_{\tau}$.
In the small $\tau$ regime, Fig. \ref{fig:3} suggests that at finite
truncations, $\beta_{d}$ is identical to that of $N_{b}=2$, namely
$\beta_{d}=(1-s)/(2s)$.

\begin{figure}
\resizebox{0.52\textwidth}{!}{
  \includegraphics{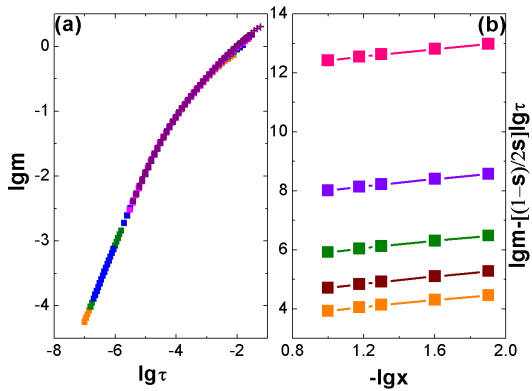}}
\caption{(a) Overlap of curves with $N_{b}=15, 20, 30, 40, 80$ to
the curve with $N_{b}=10$ after proper translation at $s=0.3$. (b)
Scaling of the magnetization with respective to the truncation. From
top to bottom $s=0.1, 0.15, 0.2, 0.25, 0.3$, respectively. Here, we
use lg$\tau$= -5.0 for $s=0.1$ and lg$\tau$=-6.0 for the others.
Parameters are $\Delta$=0.1, $\epsilon$=0.0.} \label{fig:4}
\end{figure}

Our assumptions in Eq.(\ref{eq:34}) are verified in Fig.
\ref{fig:4}(a). All the curves overlap with the curve at $N_{b}=10$
after proper translations of $\overline{m}$ and $\overline{\tau}$.
Translation details are in Tab. \ref{tab:1}. We get
$\Delta\overline{m}/\Delta \overline{\tau}=a_{m}/a_{\tau}\simeq
0.5$, supporting $\beta_{u}=\beta_{MF}=1/2$, consistent with the
specific case $N_{b}=\infty$. According to Eq.(\ref{eq:34}), the
crossover point $\tau_{cr}$ between the two power law regime
$\beta_{u}=1/2$ (upper power) and $\beta_{d}=(1-s)/(2s)$ (lower
power) is given by $\tau_{cr}\sim x^{a_{\tau}/a_{x}}$. $\tau_{cr}$
declines as $N_{b}$ increases, leading to the expansion of the
$\beta_{u}=1/2$ regime. In the limit $N_{b}=\infty$, i.e., $x=0$,
$\tau_{cr}$ is moved to zero, and the full mean field curve should
be recovered. This is indeed observed in Fig. \ref{fig:3}.

\begin{table}
\caption{Translation of $\overline{m}$, $\overline{\tau}$ and
$\overline{x}$ data to $N_{b}=10$ curve for $s=0.3$.} \label{tab:1}
\begin{tabular}{lllllll}
\hline\noalign{\smallskip}
$N_{b}$ & $\Delta\overline{x}$ & $\Delta\overline{\tau}$ & $\Delta\overline{m}$ & $a_{m}/a_{x}$ & $a_{\tau}/a_{x}$ & $a_{m}/a_{\tau}$ \\
\noalign{\smallskip}\hline\noalign{\smallskip}
15 & 0.176 & 0.20 & 0.10 & 0.57& 1.14& 0.50\\
20 & 0.301 & 0.30 & 0.15 & 0.50& 1.00& 0.50\\
30 & 0.477 & 0.48 & 0.24 & 0.50& 1.00& 0.50\\
40 & 0.602 & 0.60 & 0.29 & 0.48& 1.00& 0.48\\
80 & 0.903 & 0.86 & 0.43 & 0.48& 0.95& 0.50\\
\noalign{\smallskip}\hline
\end{tabular}
\end{table}

   In the regime $\tau<\tau_{cr}$, the second equation in
Eq.(\ref{eq:34}) holds. Fig. \ref{fig:4}(b) shows
lg$m-\beta_{d}$lg$\tau$ v.s. lg$x$ at fixed $\tau$ in the lower
regime. Here $\beta_{d}=(1-s)/(2s)$ is used. For various $s$ values,
the curves are linear with the same slope $C$. This confirms the
second equation of Eq.(\ref{eq:34}) and gives
$\beta_{d}a_{\tau}/a_{x}-a_{m}/a_{x}=C$, being independent of $s$.
Numerical fitting gives the slope $C\sim0.61\pm0.02$. Taking into
account of $a_{m}/a_{\tau}=1/2$ and $\beta_{d}=(1-s)/(2s)$, we get

\begin{eqnarray}\label{eq:35}
\frac{a_{\tau}}{a_{x}}&=&\frac{2Cs}{1-2s},
\end{eqnarray}

\begin{eqnarray}\label{eq:36}
\frac{a_{m}}{a_{x}}&=&\frac{Cs}{1-2s}.
\end{eqnarray}

\begin{figure}
\resizebox{0.52\textwidth}{!}{
  \includegraphics{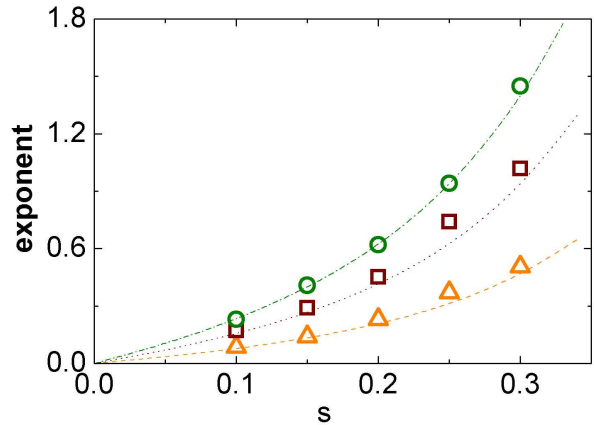}}
\caption{Ratio of the scaling powers as functions of $s$ at
$\Delta=0.1$. Orange triangles are
$a_{m}/a_{x}=\Delta\overline{m}/\Delta \overline{x}$ obtained from
translation in lg$m$ v.s. lg$\tau$ figures at different $s$'s as
Fig. (\ref{fig:4}.a) for $s=0.3$; Dash line is the function
$a_{m}/a_{x}=-0.61s/(2s-1)$. Wine squares are
$a_{\tau}/a_{x}=\Delta\overline{\tau}/\Delta \overline{x}$ from
translation in lg$m$ v.s. lg$\tau$ figures at different $s$'s like
Fig. (\ref{fig:4}.a); The dot line is $-1.22s/(2s-1)$. Olive circles
are $a_{\epsilon}/a_{x}$ from translation in lg$m$ v.s. lg$\epsilon$
figures at different $s$'s as Fig. (\ref{fig:6}.b) for $s=0.2$; The
dash-dot line is $-1.83s/(2s-1)$. }\label{fig:5}
\end{figure}

They are plotted as lines in Fig. \ref{fig:5}. On the other hand,
the average values of $a_{m}/a_{x}$ and $a_{\tau}/a_{x}$ obtained
from translation of $\overline{m}$ data are plotted in Fig.
\ref{fig:5} as functions of $s$. They match Eq.(\ref{eq:35}) and
Eq.(\ref{eq:36}) remarkably well.

   Fig.(\ref{fig:1})-(\ref{fig:4}) show that even for a mean-field
Hamiltonian, local boson state truncation can dramatically alter the
critical exponents, from the Gaussian exponents to the interacting
ones. The boson state truncation, i.e., the artificial constraint of
the local Hilbert space can be regarded as an additional local
interaction introduced between bosons. For the mean-field model with
Gaussian critical fixed point, this interaction becomes dominant in
low energies and leads the system to a new interacting critical
fixed point. The critical exponent does not change continuously with
the Hilbert space constraint. Instead, the constraint tunes the
crossover point in the $m-\tau$ curve below which the system flows
into a truncation-dominated interacting fixed point.

Our results shed some lights on the problem of bosonic NRG for the
spin-boson model. That $\beta_{d}=(1-s)/(2s)$ of $H_{MF}^{star}$
agrees with the $\beta$ obtained from NRG for the spin-boson model
suggests that $\beta_{d}$ itself is totally dominated by truncation.
It is probable that the same scenario of $N_{b}$ also occurs in NRG.
If this is true, it becomes clear why the bosonic NRG with boson
state truncation cannot give correct classical exponent in this
regime: the Gaussian critical fixed point in the regime $0<s<1/2$ is
overtaken by the strong relevant operator introduced by the Hilbert
space truncation \cite{RefVT:16,RefVTB:24}. Moreover, one cannot
improve the NRG critical exponent simply by increasing $N_{b}$ if
one focuses only on the small $\tau$ limit. Instead, the correct
exponents can be crudely observed at finite energy scales from the
truncated NRG calculation as shown in Ref.\cite{RefVTB:24}. As
demonstrated here, to fully disclose the role of $N_{b}$ and to
extract the accurate exponent $\beta$ from the bosonic NRG
calculation, it is necessary to carry out a scaling analysis of
$N_{b}$. It is worth mentioning that the NRG study of the Ising-BFKM
gives the interacting critical fixed point in the regime $0<s<1/2$,
in contrast to the belief that the Ising-BFKM and SBM belong to the
same universality class \cite{RefTI:26,RefMK:27}. In light of our
study, we suggest that the NRG study for Ising-BFKM should be
checked, using similar scaling analysis of $N_{b}$.

\begin{figure}
\resizebox{0.52\textwidth}{!}{
  \includegraphics{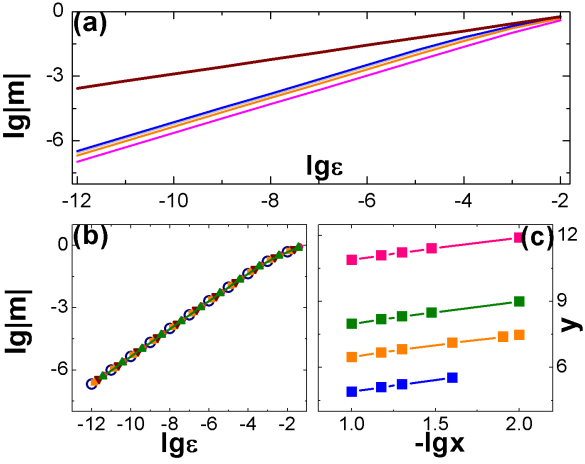}}
\caption{Magnetization $m$ as functions of bias $\epsilon$
  for different truncations at the
parameters: $\Delta=0.1$, $\alpha=\alpha_{c}$, $s=0.2$. (a)
$N_{b}=2, 10, 40, 100, \infty$ from bottom to top. (b) Overlap of
curves with $N_{b}=15, 20, 40, 80, 100$ to the curve with $N_{b}=10$
after proportional translation. (c) Scaling of the magnetization in
terms of the truncation. $s=0.1,0.15,0.2,0.3$ from top to bottom.
$y=[\overline{m}-\overline{\epsilon}(1-s)/(1+s)]/[2Cs/(1+s)]$, here
we use $\overline{\epsilon}=-9.0$ and $C=0.61$.}\label{fig:6}
\end{figure}

   In the following, we study the influence of truncation on $\delta$.
In Fig. \ref{fig:6}, we show $m$ as a function of bias $\epsilon$ at
the critical $\alpha_c$'s with different $N_{b}$'s and the collapse
of these curves to the curve at $N_{b}=10$. Fig. \ref{fig:6}(a)
shows that in the lower regime the magnetization has the same power
dependence on $\epsilon$ as that of $N_{b}=2$, i.e.,
$\delta_{d}=(1+s)/(1-s)$ as given in Eq.(\ref{eq:29}). Fig.
\ref{fig:6}(b) shows the collapse of the lg$m$-lg$\epsilon$ curves
under proper translations of $\overline{x}$, $\overline{\epsilon}$
and $\overline{m}$. This supports that, near the criticality
$m(\epsilon, x)$ is also a GHF:

\begin{eqnarray}\label{eq:37}
m(\lambda^{a_{\epsilon}}\epsilon, \lambda^{a_{x}}x)&=&
\lambda^{a_{m}}m(\epsilon, x).
\end{eqnarray}

The ratios between the shifts are $\Delta\overline{m}/\Delta
\overline{\epsilon}=a_{m}/a_{\epsilon}$ and
$\Delta\overline{m}/\Delta \overline{x}=a_{m}/a_{x}$. Details of the
translation are in Tab. \ref{tab:2}. According to Fig.
\ref{fig:6}(a) and Fig. \ref{fig:6}(b), $m$ as a GHF of $x$ and
$\epsilon$ can be formulated as:

\begin{eqnarray}\label{eq:38}
\overline{m}(\overline{\epsilon},\overline{x})= \left\{
\begin{array}{lll}
\delta^{-1}_{u}\overline{\epsilon}+C_{u},\,\,\,\,\,\,&
(\epsilon^{\frac{1}{a_{\epsilon}}}\gg
cx^{\frac{1}{a_{x}}});\\
& \\
\delta^{-1}_{d}\overline{\epsilon}+\overline{x}
(\frac{a_{m}}{a_{x}}-\frac{a_{\epsilon}}{a_{x}}\delta^{-1}_{d})+C_{d},\,\,\,\,\,\,
& (\epsilon^{\frac{1}{a_{\epsilon}}}\ll cx^{\frac{1}{a_{x}}}).
\end{array} \right.
\end{eqnarray}

We get $\delta_{u}^{-1}=a_{m}/a_{\epsilon}=\Delta\overline{m}/\Delta
\overline{\tau}\simeq 1/3$, signaling a classical power law in the
large $\epsilon$ regime. In Fig. \ref{fig:6}(c) we plot
$(\overline{m}-\delta_{d}^{-1}\overline{\epsilon})/(a_{m}/a_{x}-\delta_{d}^{-1}a_{\epsilon}/a_{x})$
versus $\overline{x}$ for different $s$. They are linear functions
of $\overline{x}$ with an average slope $1.01\pm0.05$, as expected
from the second equation of Eq.(\ref{eq:38}).

In Fig.\ref{fig:5}, $a_{\epsilon}/a_{x}$ obtained from the
translation procedure is plotted as circles, to be compared with the
analytical curve $a_{\epsilon}/a_{x}=3Cs/(1-2s)$. Here we use the
value $C=0.61$ obtained previously. They agree very well. These
numerical results confirm our assumption in Eq.(\ref{eq:38}).

\begin{table}
\caption{Translation of $\overline{m}$, $\overline{\epsilon}$ and
$\overline{x}$ data to $N_{b}=10$ curve for $s=0.2$.} \label{tab:2}
\begin{tabular}{lllllll}
\hline\noalign{\smallskip}
$N_{b}$ & $\Delta\overline{x}$ & $\Delta\overline{\epsilon}$ & $\Delta\overline{m}$ & $a_{m}/a_{x}$ & $a_{\epsilon}/a_{x}$ & $a_{m}/a_{\epsilon}$ \\
\noalign{\smallskip}\hline\noalign{\smallskip}
15 & 0.176 & 0.11 & 0.036 & 0.20 & 0.63 & 0.33\\
20 & 0.301 & 0.19 & 0.063 & 0.21 & 0.63 & 0.33\\
40 & 0.602 & 0.36 & 0.12 & 0.20 & 0.60 & 0.33\\
80 & 0.903 & 0.56 & 0.19 & 0.21 & 0.62 & 0.34\\
100& 1.000 & 0.63 & 0.21 & 0.21 & 0.63 & 0.33\\
\noalign{\smallskip}\hline
\end{tabular}
\end{table}

This shows that $m(\epsilon, x)$ is also a function with two
different power law regimes: the low energy regime with
$1/\delta_{d}$ and the upper one with $1/\delta_{MF}=1/3$. They are
separated by a crossover point $\epsilon_{cr}\sim
x^{a_{\epsilon}/a_{x}}$, similar to the situation in $m(\tau, x)$.

\begin{figure}
\resizebox{0.52\textwidth}{!}{
  \includegraphics{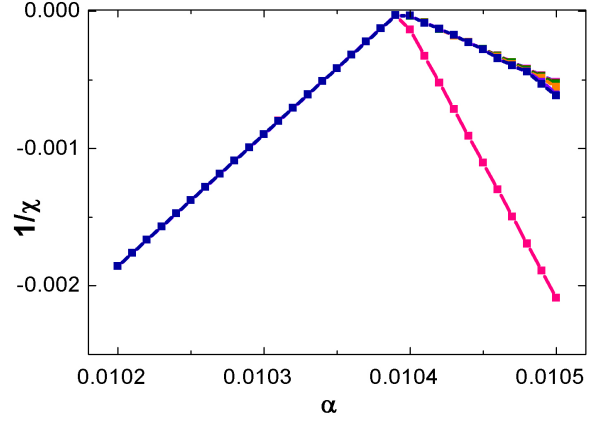}}
\caption{Inverse of susceptibility $1/\chi$ as a function of
$\alpha$ for different truncations $N_{b}$ at the parameters:
$\Delta=0.1$, s=0.2, $\epsilon=0$. The zero point of $1/\chi$ gives
the critical coupling strength $\alpha_{c}$. $N_{b}=2$, 5, 10, 15,
20, 40, 80, 100, $\infty$ from top to bottom,
respectively.}\label{fig:7}
\end{figure}

   In the following, we discuss the susceptibility exponent
$\gamma$. In Fig. \ref{fig:7}, $1/\chi$ versus $\alpha$ is plotted.
The zero point in the figure is the critical point. Within the
critical region, the susceptibility exponent remains the classical
one, $\gamma=1$ under finite boson state truncation in $0<s<1/2$.
This shows that the boson state truncation does not influence
$\gamma$. Similar observations are made in the NRG, MC, SPSA and
extended coherent state approach studies
\cite{RefVT:16,RefWR:18,RefAAH:19,RefYQK:28}.

   From the analysis above, we observe that in the low energy regimes, i.e.,
$\tau^{1/a_{\tau}}\ll cx^{1/a_{x}}$ and
$\epsilon^{1/a_{\epsilon}}\ll cx^{1/a_{x}}$, $\beta$ and $\delta$
are no longer determined by the scaling exponents $a_{m}$,
$a_{\tau}$, $a_{\epsilon}$. Instead new parameters are needed to
define those critical exponents. We also find that in both the high
and low energy regimes, the scaling relation
$\beta(\delta-1)=\gamma$ is fulfilled.

   The same analysis is carried out at a different $\Delta$ and
$\Lambda$: $\Delta=0.2$ and $\Lambda=3.0$. Conclusions are
qualitatively the same. From the analysis above, we believe that it
is not the failure of quantum-to-classical mapping for the
spin-boson model, but the numerical concomitant, i.e., the
truncation that leads to new exponents in the regime $0<s<1/2$. Our
work shows that the boson state truncation $N_{b}$ severely changes
the critical behavior of the mean-field spin-boson model: it plays
the role of a new scaling parameter and changes the low energy
exponent $\beta$ and $\delta$ into a truncation-dominated exponent
(Eq.(\ref{eq:27}) and Eq.(\ref{eq:29})).

\section{Summary}
\label{sec:8} To conclude, we have analyzed the mean-field spin
boson model under boson state truncation. We focus on the quantum
critical behavior of the model and carry out the scaling analysis of
the magnetization in terms of truncation parameter $N_{b}$. Our work
shows that for the mean field spin boson model the truncation gives
rise to a relevant perturbation in the regime $0<s<1/2$. The effect
of the truncation, while cutting no ice for the exponents $\beta$
and $\delta$ in the regime $s>1/2$, dominates in $0<s<1/2$. As is
shown in in Fig. \ref{fig:3} and in Fig. \ref{fig:6}, the
magnetization $m$ under truncation is a generalized homogeneous
function in the regime $0<s<1/2$: in the high energy regime, $m$
fulfills a classical power law; in the low energy regime it exhibits
a double power behavior, and the truncation $N_{b}$ becomes a new
scaling parameter, besides the coupling strength $\tau$ and external
field $\epsilon$ (Eq.(\ref{eq:34}) and Eq.(\ref{eq:38})). Moreover,
we find that as $N_{b}$ increases, the crossover point between two
power law regimes moves downwards and to zero energy scale at
$N_{b}=\infty$ (namely without boson states truncation).
Implications of these findings to the bosonic NRG study are
discussed. For the spin-boson model within $0<s<1/2$, it is now
believed that the low energy critical exponents $\beta=(1-s)/(2s)$
and $\delta=(1+s)/(1-s)$ obtained by NRG are also due to boson state
truncation $N_{b}$ \cite{RefVT:16,RefVTB:24}. We conjecture that
similar scaling behavior of $N_{b}$ should also occur in the NRG
study of the spin-boson model. A correct and accurate extraction of
$\beta$ and $\delta$ for the spin-boson model from NRG requires
similar scaling analysis as done here. Further studies in this
direction are in progress.

   N. H. Tong acknowledges helpful discussions with Ralf Bulla and
Matthias Vojta. This work is supported by National Basic Research
Program of China (grant number 2007CB925004), and by the NSFC under
grant number 10674178.

\appendix{}
\section{Appendix}
In the Appendix we give the analytical derivation of the critical
exponents at $N_{b}=2$. Following similar procedures in Sec. 2.2, we
can get the self-consistent equations for the magnetization m at
$N_{b}=2$

\begin{eqnarray}\label{eq:39}
m&=&\frac{\Delta^{2}-(\epsilon +\epsilon
^{\prime}+t)^{2}}{\Delta^{2}+ (\epsilon +\epsilon ^{\prime}+t)^{2}},
\end{eqnarray}

\begin{eqnarray}\label{eq:40}
t&=&\sqrt{(\epsilon +\epsilon ^{\prime})^{2}+\Delta^{2}},
\end{eqnarray}

\begin{eqnarray}\label{eq:41}
\epsilon ^{\prime}&=&-\frac{m}{\pi}\sum_{n}d_{n},
\end{eqnarray}

\begin{eqnarray}\label{eq:42}
d_{n}&=&\frac{\gamma_{n}^{2}}{\sqrt{\xi_{n}^{2}+\frac{m^{2}\gamma_{n}^2}{\pi}}}.
\end{eqnarray}
with $\gamma_{n}$ and $\xi_{n}$ satisfying Eq.(\ref{eq:11}) and
Eq.(\ref{eq:12}), respectively. Since we are interested in the
critical exponent $\beta$, first we take the bias $\epsilon $ to be
zero. Combining Eq.(\ref{eq:39}) and Eq.(\ref{eq:40}), we get the
following equation:

\begin{eqnarray}\label{eq:43}
m&=&\frac{\epsilon ^{\prime}}{ \sqrt{{\epsilon
^{\prime}}^{2}+\Delta^{2}}}.
\end{eqnarray}

The newly obtained Eq.(\ref{eq:43}) together with Eq.(\ref{eq:41})
composes the self-consistent equations. To calculate the summation
in Eq.(\ref{eq:41}), we define a parameter $n_{0}$ determined by the
following equation:

\begin{eqnarray}\label{eq:44}
\xi_{n_{0}}^{2}=\frac{m^{2}\gamma_{n_{0}}^2}{\pi}.
\end{eqnarray}

Then Eq.(\ref{eq:41}) can be divided into two parts:

\begin{eqnarray}\label{eq:45}
\epsilon
^{\prime}&=&-\frac{m}{\pi}[{\sum_{n=0}^{n_{0}}d_{n}+\sum_{n=n_{0}+1}^{\infty}d_{n}}].
\end{eqnarray}

When $n\ll n_{0}$, i.e.,
$\xi_{n}^{2}\gg\frac{m^{2}\gamma_{n}^2}{\pi}$,

\begin{eqnarray}\label{eq:46}
d_{n}&=&\frac{\gamma_{n}^{2}}{\xi_{n}}(1+\frac{m^{2}\gamma_{n}^{2}}{\pi\xi_{n}^{2}})^{-\frac{1}{2}}
\nonumber \\&=&\sum_{k=0}^{\infty}P_{k}\Lambda^{n[k(1-s)-s]}m^{2k}.
\end{eqnarray}

On the contrary,when $n\gg n_{0}$, i.e.,
$\xi_{n}^{2}\ll\frac{m^{2}\gamma_{n}^2}{\pi}$,

\begin{eqnarray}\label{eq:47}
d_{n}&=&\frac{\sqrt{\pi}\gamma_{n}}{m}
(1+\frac{\xi_{n}^{2}\pi}{m^2\gamma_{n}^{2}}) ^{-\frac{1}{2}}
\nonumber
\\&=&\sum_{k=0}^{\infty}Q_{k}\Lambda^{\frac{n[2k(s-1)-1-s]}{2}}m^{-2k-1}.
\end{eqnarray}

Here, $P_{k}=\frac{C_{k}^{-1/2}}{\pi^{k}}\frac{A^{2k+2}}{B^{2k+1}}$
and $Q_{k}=\frac{C_{k}^{-1/2}\pi^{k+1/2}B^{2k}}{A^{2k-1}}$, with
$C_{k}^{-1/2}=-\frac{1}{2}(-\frac{1}{2}-1)...[-\frac{1}{2}-(k-1)]$
($C_{0}^{-1/2}=1$), $A^{2}=\gamma_{n}^{2}\Lambda^{n(1+s)}$ and
$B=\xi_{n}\Lambda^{n}$. Replacing $d_{n}$ in Eq.(\ref{eq:45}) with
Eq.(\ref{eq:46}) and Eq.(\ref{eq:47}), respectively, we arrive at
the summation:

\begin{eqnarray}\label{eq:48}
\epsilon^{\prime}&=&-\frac{m}{\pi}[\sum_{k}P_{k}^{\prime}m^{2k}-Tm^{\frac{2s}{1-s}}].
\end{eqnarray}

Here, $P_{k}^{\prime}=P_{k}/[1-\Lambda^{k(1-s)-s}]$. It is hard to
get the analytical form for $T$, anyhow, it does not contribute to
the exponent of the order parameter. Neglecting the trivial solution
$m=0$, we get the following self-consistent equation by substituting
$\epsilon ^{\prime}$ in Eq.(\ref{eq:43}) with Eq.(\ref{eq:48}),

\begin{eqnarray}\label{eq:49}
[\sum_{k}P_{k}^{\prime}m^{2k}-Tm^{\frac{2s}{1-s}}]^{2}(1-m^{2})&=&\pi^{2}\Delta^{2}.
\end{eqnarray}

Considering that the order parameter approaches zero near the
critical coupling $\alpha_{c}$, we get

\begin{eqnarray}\label{eq:50}
P_{0}^{\prime}+P_{1}^{\prime}m^{2}-Tm^{\frac{2s}{1-s}}&=&\pi\Delta.
\end{eqnarray}

The final critical coupling strength $\alpha_{c}$ is

\begin{eqnarray}\label{eq:51}
\alpha_{c}&=&\frac{\Delta(s+1)^{2}(1-\Lambda^{-s})[1-(\Lambda^{-(s+2)})]}{2\omega_{c}(s+2)[1-(\Lambda^{-(s+1)})]^{2}},
\end{eqnarray}
the same as in Eq.(\ref{eq:26}) for $N_{b}=\infty$. For $s>1/2$, the
third term in Eq.(\ref{eq:50}) is negligible and one gets
$m\propto(\alpha-\alpha_{c})^{1/2}$; for $s=1/2$, the second and
third term have the same power, i.e., $(1-s)/(2s)=1/2$ (s=1/2), so
one gets $m\propto(\alpha-\alpha_{c})^{1/2}$; for $0<s<1/2$,
ignoring the second term in Eq.(\ref{eq:50}), one gets
$m\propto(\alpha-\alpha_{c})^{(1-s)/(2s)}$. In summary, for
$N_{b}=2$, we get

\begin{eqnarray}\label{eq:52}
\beta&=&\left\{
\begin{array}{lll}
\frac{1}{2},\,\,\,\,\,\,\,\,\, & (\textrm{$s\geq\frac{1}{2}$});\\
 &\\
 \frac{1-s}{2s},\,\,\,\,\,\,\,\,\,&
(\textrm{$0<s<\frac{1}{2}$}).
\end{array} \right.
\end{eqnarray}

   In the case of $s=0$, a small difference lies in Eq.(\ref{eq:46}) at
$k=0$. If $s=0$ and $k=0$, the common ratio of the geometric series
in Eq.(\ref{eq:45}) for $n<n_{0}$, namely $\Lambda^{[k(1-s)-s]}$, is
unity and the general summation formula is not applicable. Taking
this speciality into account and following similar procedure of the
case $s\neq0$, one gets the critical behavior of the magnetization
as

\begin{eqnarray}\label{eq:53}
m\propto\alpha^{-\frac{1}{2}}e^{-\frac{\Delta
ln\Lambda(1+\Lambda^{^{-1}})}{8\alpha\omega_{c}(1-\Lambda^{-1})}}.
\end{eqnarray}

   As far as $\delta$ is concerned, similar analysis
gives the following expression

\begin{eqnarray}\label{eq:54}
\delta&=&\left\{
\begin{array}{lll}
3,\,\,\,\,\,\,\,\,\, & (\textrm{$s\geq\frac{1}{2}$});\\
 &\\
 \frac{1+s}{1-s},\,\,\,\,\,\,\,\,\,&
(\textrm{$0<s<\frac{1}{2}$}).
\end{array} \right.
\end{eqnarray}

\end{document}